\documentclass[reprint, twocolumn, superscriptaddress,prb]{revtex4-2}
\usepackage{bm, amsmath, amsfonts, amssymb, braket}
\usepackage{subfigure}
\usepackage{color}
\usepackage{graphicx}
\usepackage{dcolumn} 

\usepackage{rotating} 

\begin{document}
\title{Boundary-dependent dynamical instability of bosonic Green's function:
Dissipative Bogoliubov-de Gennes Hamiltonian and its application to non-Hermitian skin effect}
\author{Nobuyuki Okuma}
\email{okuma@hosi.phys.s.u-tokyo.ac.jp}
\affiliation{%
 Yukawa Institute for Theoretical Physics, Kyoto University, Kyoto 606-8502, Japan
}%

\date{\today}
\begin{abstract}
The energy spectrum of bosonic excitations from a condensate is given by the spectrum of a non-Hermitian Hamiltonian constructed from a bosonic Bogoliubov-de Gennes (BdG) Hamiltonian in general even though the system is essentially Hermitian.
In other words, two types of non-Hermiticity can coexist: one from the bosonic BdG nature and the other from the open quantum nature.
In this paper, we propose boundary-dependent dynamical instability.
We first define the bosonic dissipative BdG Hamiltonian in terms of Green's function in Nambu space and discuss the correct particle-hole symmetry of the corresponding non-Hermitian Hamiltonian.
We then construct a model of the boundary-dependent dynamical instability so that it satisfies the correct particle-hole symmetry.
In this model, an anomalous term that breaks the particle number conservation represents the non-Hermiticity of the BdG nature, while a normal term is given by a dissipative Hatano-Nelson model.
Thanks to the competition between the two types of non-Hermiticity, the imaginary part of the spectrum can be positive without the help of the amplification of the normal part and the particle-hole band touching that causes the Landau instability.
This leads to the boundary-dependent dynamical instability
under the non-Hermitian skin effect, -strong dependence of spectra on boundary conditions for non-Hermitian Hamiltonians-,
of the Bogoliubov spectrum.

\end{abstract}
\maketitle
\section{Introduction}
At low temperature, the excitations from a Bose-Einstein condensate are well described by a bosonic Bogoliubov-de Gennes (BdG) Hamiltonian \cite{bogoliubov1947theory,Altland-Simons,Colpa-78,kawaguchi2012spinor}.
As well as the systems of boson particles such as cold atoms \cite{kawaguchi2012spinor} and photons \cite{ozawa2019topological}, the BdG Hamiltonian has been extensively studied in the systems of emergent bosonic excitations from ordered states such as the magnons \cite{topo-magnon} and phonons \cite{topologicalphonon}.
Remarkably, the excitation spectrum of a bosonic BdG system is given by the eigenspectrum of a non-Hermitian matrix constructed from the bosonic BdG Hamiltonian, in the presence of $anomalous$ terms violating the particle number conservation \cite{bogoliubov1947theory,Colpa-78,kawaguchi2012spinor,Altland-Simons}.
In this sense, the bosonic BdG Hamiltonian has been interested in terms of non-Hermitian physics \cite{Bender-98, Bender-02, Bender-review, Konotop-review, Christodoulides-review,ashida2020non}, even when the system is essentially Hermitian.
In other words, there can coexist two types of non-Hermiticity in bosonic systems: one from the bosonic BdG nature and the other from the open quantum nature.

In this paper, we consider dissipative bosonic systems under the non-Hermitian skin effect \cite{YW-18-SSH}, which is the extreme sensitivity of the non-Hermitian spectrum to the boundary condition.
For this sake, we define the dissipative BdG Hamiltonian in terms of the Green's function \cite{Fetter} and determine the symmetry of it.
While one can define two equivalent particle-hole symmetries in general non-dissipative BdG Hamiltonians, we find that the natural extension to the Green's function corresponds to only one of them.
By noticing this symmetry, we construct a dissipative bosonic model in which the normal part is given by the dissipative system of the non-Hermitian skin effect.
Although the non-Hermiticity in the normal part does not correspond to an amplification, this model shows a dynamical instability \cite{kawaguchi2012spinor} that occurs under the periodic boundary condition (PBC) and does not occur under the open boundary condition (OBC) for some parameter region, without the help of a band touching between the particle and hole bands of Bogoliubov excitations.
This result can be understood in terms of a competition between an anomalous term representing the BdG non-Hermiticity and the skin effect representing the open quantum nature.

This paper is organized as follows.
In Sect.~\ref{bdg}, we define the frequency-dependent dissipative BdG Hamiltonian in terms of the Green's function in Nambu space.
We find that the correct particle-hole symmetry is given by a conjugate-type symmetry in this formalism.
In Sect.~\ref{dynamicalskin}, we propose a dynamical instability that occurs under the PBC and does not occur under the OBC, without the help of the band touching between the particle and hole bands.
We construct a dissipative BdG Hamiltonian that shows the boundary-dependent dynamical instability and analyze the complex Bogoliubov spectrum.
We also discuss the dynamical instabilities induced by infinitesimal perturbations. 
In Sect.~\ref{discussion}, we point out some subtle points about the Landau instability in terms of the density of states.

\section{Formalism\label{bdg}}
In this section, we define the dissipative Bogoliubov-de Gennes Hamiltonian.
We first review the free bosonic case and discuss two types of non-Hermitian particle-hole symmetry: the conjugate-type and the transpose-type symmetries, which are identical in the absence of dissipation.
We then define the frequency-dependent BdG Hamiltonian in terms of the Green's function in Nambu space and discuss the complex energy spectrum.
In this formalism, the correct particle-hole symmetry is given by the conjugate-type symmetry.

\subsection{Bosonic Bogoliubov-de Gennes Hamiltonian without dissipation}
Before introducing the self-energy, 
we consider free bosonic Hamiltonians, i.e.,  quadratic Hamiltonians of bosonic annihilation/creation operators ($a_i,a_i^{\dagger}$) with $i$ being the quantum degrees of freedom such as the site index and the orbital.
In many applications, $anomalous$ terms that violate the particle number conservation can exist.
In such a case, the corresponding free Hamiltonian is called the bosonic Bogoliubov-de Gennes (BdG) Hamiltonian and written in terms of the Nambu spinor $\bm{\Phi}=(\bm{a},\bm{a}^{\dagger})$:
\begin{align}
    \hat{H}=\bm{\Phi}^{\dagger}H^{(0)}_{\rm BdG}\bm{\Phi}=
    (\bm{a}^{\dagger},\bm{a}) H^{(0)}_{\rm BdG}
    \begin{pmatrix}
        \bm{a}\\
        \bm{a}^{\dagger}
    \end{pmatrix}.\label{freebdg}
\end{align}
In general, the bosonic BdG Hamiltonian matrix has the following form:
\begin{align}
    H^{(0)}_{\rm BdG}=
    \begin{pmatrix}
    h&s\\
    s^{\dagger}&h^T
    \end{pmatrix}=
    \begin{pmatrix}
    h&s\\
    s^*&h^*
    \end{pmatrix},
\end{align}
where $h$ and $s$ are Hermitian and symmetric matrices, respectively.
We have assumed the Hermiticity of $H^{(0)}_{\rm BdG}$.
Owing to this constraint, one can define the bosonic analogy of particle-hole symmetry \cite{BdGsym}:
\begin{align}
    \sigma_x[H^{(0)}_{\rm BdG}]^*\sigma_x=H^{(0)}_{\rm BdG},\label{bosonphs}
\end{align}
where $\sigma$ denotes the Pauli matrices in Nambu space henceforth.

Unlike a conventional quadratic Hamiltonian, one-particle energy spectrum of Eq. (\ref{freebdg}) is not given by the eigenvalues of the matrix representation $H^{(0)}_{\rm BdG}$ in the presence of the anomalous terms.
This is because the unitary matrix that diagonalizes $H^{(0)}_{\rm BdG}$ can ruin the bosonic commutation.
In general, one-particle energy spectrum is given by the bosonic Bogoliubov transformation \cite{Colpa-78,Gingras,Shindou-13,kawaguchi2012spinor}.
We here introduce the mapping to the fermionic non-Hermitian Hamiltonin:
\begin{align}
    H^{(0)}_{\sigma\rm BdG}=\sigma_z H^{(0)}_{\rm BdG}.
\end{align}
Corresponding to the bosonic particle-hole symmetry (\ref{bosonphs}), this non-Hermitian Hamiltonian has the fermionic particle-hole symmetry \cite{BdGsym}:
\begin{align}
    \sigma_x[H^{(0)}_{\sigma\rm BdG}]^*\sigma_x=-H^{(0)}_{\sigma\rm BdG}.\label{fermionphs}
\end{align}
In addition, the Hermiticity of $H^{(0)}_{\rm BdG}$ leads to the pseudo-Hermiticity of $H^{(0)}_{\sigma\rm BdG}$ \cite{BdGsym}:
\begin{align}
    \sigma_z[H^{(0)}_{\sigma\rm BdG}]^{\dagger}\sigma_z=H^{(0)}_{\sigma\rm BdG}.
\end{align}
Actually, the eigenvalue problem of $H^{(0)}_{\sigma\rm BdG}$ 
is directly related to the bosonic excitation spectrum:
\begin{align}
    P^{-1}H^{(0)}_{\sigma\rm BdG}P=\frac{1}{2}
    \begin{pmatrix}
    \Lambda&0\\
    0&-\Lambda
    \end{pmatrix},
\end{align}
where $\Lambda$ is a diagonal matrix whose elements give the one-particle energy spectrum of Eq. (\ref{freebdg}).
In particular, $P$ can be taken as a paraunitary matrix if the spectrum of $H^{(0)}_{\sigma\rm BdG}$ is real \cite{KSUS-19}:
\begin{align}
    P\sigma_zP^{\dagger}=\sigma_z.
\end{align}
In this case, the particle (hole) eigenstates of $H^{(0)}_{\sigma\rm BdG}$ are also eigenstates of $\sigma_z$ with an eigenvalue $+1~(-1)$ \cite{KSUS-19}.
Using the paraunitary condition, the bosonic Bogoliubov transformation is written in terms of the bosonic BdG Hamiltonian:
\begin{align}
    P^{\dagger}H^{(0)}_{\rm BdG}P=\frac{1}{2}
    \begin{pmatrix}
    \Lambda&0\\
    0&\Lambda
    \end{pmatrix}.
\end{align}
Note that the presence of the anomalous terms changes the Fock vacuum of the eigenvectors from that of $a_i$'s.
This change of the vacuum is called quantum correction. 

In most applications to solid-state physics, the (semi) positive-definiteness of $H^{(0)}_{\rm BdG}$ is further assumed, which ensures nonnegative excitation energies.
The instability that comes from the presence of negative excitation energy is called the Landau instability \cite{kawaguchi2012spinor}.
In isolated quantum systems, on the other hand, Landau instability is suppressed, and this assumption is not always imposed.
Such a situation is thought to be realized in cold atom systems.
In the absence of the positive definiteness of $H^{(0)}_{\rm BdG}$, the excitation energy can be negative or complex.
Owing to the pseudo-Hermiticity, the presence of a complex energy is equivalent to the presence of an energy with positive imaginary part.
Such an excitation mode grows in time and leads to the instability of the vacuum. This type of the instability is called dynamical instability \cite{kawaguchi2012spinor}.

Note that the following particle-hole symmetry is also valid in the presence of the pseudo-Hermiticity \cite{KSUS-19}:
\begin{align}
    \sigma_y [H^{(0)}_{\sigma\rm BdG}]^{T}\sigma_y=-H^{(0)}_{\sigma\rm BdG}.\label{transposesym}
\end{align}
In the absence of the pseudo-Hermiticity that comes from the Hermiticity of $\hat{H}$, however, the symmetries (\ref{fermionphs}) and (\ref{transposesym}) are not identical. The correct particle-hole symmetry may depend on the situation.
As we will discuss later, a natural extension of Eq. (\ref{fermionphs}) is the correct particle-hole symmetry for the effective Hamiltonian defined in terms of the Green's function.

\subsection{Bosonic Green's function in Nambu representation}
There are several ways to treat open quantum systems.
A typical mathematical tool is the Lindblad equation \cite{Lindblad, Prosen-2008}.
For example, a bosonic quadratic Lindblad equation has been studied in terms of topology and skin effect \cite{Flynn-21}. 
Another useful mathematical tool is the Green's function \cite{Fetter,keldysh1965diagram}.
Several studies have focused on the relation between the non-Hermitian skin effect and the Green's functions \cite{Borgnia-19,Okuma-correlated,Federico}. 

In the following, we consider the retarded/advanced Green's function for bosonic excitations to include the dissipation/interaction/disorder effects.
In the presence of the anomalous term known as $anomalous$ $Green's$ $function$, it is convenient to introduce the Nambu representation, as in the case of the previous subsection.
The retarded Green's function in Nambu representation is defined in terms of the Heisenberg operators \cite{Fetter}:
\begin{align}
    G_{ij}^R(t,t')&=-i\theta(t-t')\langle[\Phi_i(t),\Phi_j^{\dagger}(t')]\rangle\notag\\
    &=-i\theta(t-t')
    \begin{pmatrix}
    \langle[a_i(t),a_j^{\dagger}(t')]\rangle&\langle[a_i(t),a_j(t')]\rangle\\
    \langle[a^{\dagger}_i(t),a^{\dagger}_j(t')]\rangle&\langle[a^{\dagger}_i(t),a_j(t')]\rangle
    \end{pmatrix}\notag\\
    &=:
    \begin{pmatrix}
    A^R_{ij}(t,t')&B^R_{ij}(t,t')\\
    C^R_{ij}(t,t')&D^R_{ij}(t,t')
    \end{pmatrix},
\end{align}
where $\langle \cdot\rangle$ denotes the statistical average.
The following discussion does not depend on the choice of the statistical average, and one can choose both equilibrium \cite{Fetter} and non-equilibrium statics \cite{keldysh1965diagram,rammer2011quantum}.
Similarly, the advanced Green's function is defined as
\begin{align}
    G_{ij}^A(t,t')&=i\theta(t'-t)\langle[\Phi_i(t),\Phi_j^{\dagger}(t')]\rangle\notag\\
    &=i\theta(t'-t)
    \begin{pmatrix}
    \langle[a_i(t),a_j^{\dagger}(t')]\rangle&\langle[a_i(t),a_j(t')]\rangle\\
    \langle[a^{\dagger}_i(t),a^{\dagger}_j(t')]\rangle&\langle[a^{\dagger}_i(t),a_j(t')]\rangle
    \end{pmatrix}\notag\\
    &=:
    \begin{pmatrix}
    A^A_{ij}(t,t')&B^A_{ij}(t,t')\\
    C^A_{ij}(t,t')&D^A_{ij}(t,t')
    \end{pmatrix}.
\end{align}

In the following, we derive relationships between the matrices $A^{R/A},B^{R/A},C^{R/A},$ and $D^{R/A}$.
By definition, the following relationships hold:
\begin{align}
    D^R_{ij}(t,t')&=i\theta(t-t')\langle[a_j(t'),a_i^{\dagger}(t)]\rangle=A^A_{ji}(t',t),\notag\\
    D^A_{ij}(t,t')&=-i\theta(t'-t)\langle[a_j(t'),a_i^{\dagger}(t)]\rangle=A^R_{ji}(t',t),\notag\\
    B^R_{ij}(t,t')&=i\theta(t-t')\langle[a_j(t'),a_i(t)]\rangle=B^A_{ji}(t',t),\notag\\
    C^R_{ij}(t,t')&=i\theta(t-t')\langle[a^{\dagger}_j(t'),a^{\dagger}_i(t)]\rangle=C^A_{ji}(t',t).
\end{align}
In the steady state, the Green's functions have the time translation invariance, and these relations are written in the Fourier form:
\begin{align}
    D^R(\omega)&=[A^A(-\omega)]^T,D^A(\omega)=[A^R(-\omega)]^T,\notag\\
    B^R(\omega)&=[B^A(-\omega)]^T,C^R(\omega)=[C^A(-\omega)]^T.
\end{align}
By combining these equations with $[G^R(\omega)]^{\dagger}=G^A(\omega)$,
we obtain
\begin{align}
    G^{R/A}(\omega)=
    \begin{pmatrix}
    A^{R/A}(\omega)&B^{R/A}(\omega)\\
    [B^{R/A}(-\omega)]^{*}&[A^{R/A}(-\omega)]^{*}
    \end{pmatrix}.
\end{align}
From this expression, one can find the following symmetry for the retarded/advanced Green's function:
\begin{align}
    \sigma_x[G^{R/A}(-\omega)]^*\sigma_x=G^{R/A}(\omega).\label{gfsym}
\end{align}

\subsection{Dissipative Bogoliubov-de Gennes Hamiltonian defined for Green's function}
For a free bosonic system, the Green's function in Nambu space is defined via
\begin{align}
    G^{-1}_0=\omega\sigma_z-H^{(0)}_{\rm BdG}.
\end{align}
In the presence of the self-energy part $\Sigma^R(\omega)$, the retarded Green's function is given by
\begin{align}
    G^R(\omega)&=\frac{1}{G^{-1}_0-\Sigma^R(\omega)}=\frac{1}{\omega\sigma_z-[H^{(0)}_{\rm BdG}+\Sigma^R(\omega)]}\notag\\
    &=:\frac{1}{\omega\sigma_z-H^{(\rm eff)}_{\rm BdG}(\omega)}.
\end{align}
Here we have defined the dissipative BdG Hamiltonian $H^{(\rm eff)}_{\rm BdG}(\omega)$, which is a frequency-dependent non-Hermitian matrix.
From the symmetry of the retarded Green's function (\ref{gfsym}), one can still define the bosonic particle-Hole symmetry for $H^{(\rm eff)}_{\rm BdG}(\omega)$:
\begin{align}
    \sigma_x [H^{(\rm eff)}_{\rm BdG}(-\omega)]^*\sigma_x=H^{(\rm eff)}_{\rm BdG}(\omega).
\end{align}
\subsection{Eigenvalue problem of dissipative Bogoliubov-de Gennes Hamiltonian}
The complex energy spectrum of the quasi-particles is defined by the set of poles of the retarded Green's function \cite{sieberer2016keldysh} up to the factor $2$:
\begin{align}
    0=\det [G^{R}(z)^{-1}]&=\det [\sigma_z]\det [z-\sigma_zH^{(\rm eff)}_{\rm  BdG}(z)]\notag\\&=-\det [z-H^{(\rm eff)}_{\sigma\rm  BdG}(z)],\label{poles}
\end{align}
where $G^R(z)$ is the analytic continuation of $G^R(\omega)$.
We have defined the fermionic counterpart $H^{(\rm eff)}_{\sigma\rm  BdG}=\sigma_zH^{(\rm eff)}_{\rm  BdG}$.
As in the case of the free bosons, the eigenvalues are calculated in terms of the fermionic BdG Hamiltonian.
Note that there is no longer pseudo-Hermiticity in the fermionic counterpart owing to the non-Hermiticity of $H^{(\rm eff)}_{\rm BdG}(\omega)$.
This implies that the equivalence between the conjugate-type symmetry and the transpose-type symmetry does not hold for the fermionic BdG Hamiltonian. 
In the present formalism, the proper particle-hole symmetry of the fermionic BdG Hamiltonian is given by the following conjugate-type symmetry:
\begin{align}
    \sigma_x [H^{(\rm eff)}_{\sigma\rm BdG}(-\omega)]^*\sigma_x=-H^{(\rm eff)}_{\sigma\rm BdG}(\omega).\label{omegadepphs}
\end{align}
In the absence of the frequency dependence of the self-energy, this symmetry is reduced to the symmetry (\ref{fermionphs}).
Note that we can also define the dissipative fermionic BdG Hamiltonian $H^{(\rm eff)}_{\rm fBdG}$ for the fermionic Green's function via $G^R(\omega)=[\omega-H^{(\rm eff)}_{\rm fBdG}]^{-1}$, which describes superconductors.
The same symmetry (\ref{omegadepphs}) holds for that case. The derivation is almost equal to the bosonic case except for the difference between commutation and anti-commutation relations. This symmetry was discussed in terms of the dissipative Majorana fermions \cite{beenakker2015random}.

Finally, we remark the symmetry constraints for poles of the retarded Green's function.
For an eigenvalue of $H^{(\rm eff)}_{\sigma\rm  BdG}$, $E(\omega)$, one can find the particle-hole counterpart $-E^*_{-\omega}$ owing to the particle-hole symmetry (\ref{omegadepphs}). 
Correspondingly, for a pole of retarded Green's function, $z_{\rm p}$, one can also find the particle-hole counterpart $-z_{\rm p}^*$.
This is because the followings hold for
a pair of poles:
\begin{align}
    z&=E(z),\label{pole1}\\
    z&=-[E(-z^*)]^*,\label{pole2}
\end{align}
where $E(z)$ and $-[E(-z^*)]^*$ are the analytic continuations of $E(\omega)$ and $-E^*_{-\omega}$, respectively.
If $z=z_p$ satisfies Eq. (\ref{pole1}), then $z=-z_p^*$ does Eq. (\ref{pole2}), and vice versa.

\subsection{Dynamical instability in bosonic steady state}
In bosonic systems with nonzero self-energy, the presence or absence of the dynamical instability can be predicted by using the imaginary part of the complex energy spectrum defined by Eq. (\ref{poles}).
If there is a complex energy with positive imaginary part, such an excitation grows in time and indicates the instability of the steady state that is assumed in the quadratic analysis. 

\subsection{Particle and hole bands}
The definitions of particle and hole bands are not obvious in dissipative systems. In the following, we limit our discussion to the poles of $\sigma_zG^R(z)$ that is adiabatically connected to $(z-\sigma_z)^{-1}$ without closing the real energy gap. If a pole is mapped to the pole $z=+1~(-1)$ under this process, we say it is in a particle (hole) hole band.

\section{Boundary-dependent dynamical instability under non-Hermitian skin effect\label{dynamicalskin}}
In this section, we propose a dynamical instability that occurs under the PBC and does not occur under the OBC, without the help of the band touching between the particle and hole bands.
We first review the basics of the non-Hermitian skin effect.
We then construct a dissipative BdG Hamiltonian that shows the boundary-dependent dynamical instability, by noticing the conjugate-type particle-hole symmetry of the BdG Hamiltonian.
We analyze the complex Bogoliubov spectrum and discuss the competition between two types of non-Hermiticity: one from the anomalous term and the other from the self-energy.

\subsection{Hatano-Nelson model and non-Hermitian skin effect}
We here briefly review the non-Hermitian skin effect.
Non-Hermitian skin effect is a phenomenon in which the OBC spectrum of a non-Hermitian Hamiltonian is drastically different from the PBC one \cite{YW-18-SSH}.
The simplest example is the Hatano-Nelson model without disorder \cite{Hatano-Nelson-96,Hatano-Nelson-97}:
\begin{align}
    \hat{H}^{\rm(HN)}=\sum_{i}\left[(t+g)c^\dagger_{i+1}c_i+(t-g)c^\dagger_i c_{i+1}\right],
\end{align}
where $i$ is the site index, $(c,c^{\dagger})$ are creation/annihilation operators, and $t\in\mathbb{R}$ and $g\in\mathbb{R}$ are the Hermitian symmetric and non-Hermitian asymmetric hopping terms, respectively. Regardless of the statistics of field operators, the one-particle energy is simply given by the eigenvalues of the matrix form of the Hamiltonian:
\begin{align}
H^{\rm(HN)}:=
\begin{pmatrix}
0&t-g&0&\cdots\\
t+g&0&t-g&\cdots\\
0&t+g&0&\cdots\\
\vdots&\vdots&\vdots&\ddots
\end{pmatrix}.\label{hnham}
\end{align}
By using the Fourier transform, one obtain the PBC spectrum parameterized by a crystal momentum $k$:
\begin{align}
    E^{\rm (HN,PBC)}_{k}=2t\cos k+2ig\sin k.\label{pbcspectrum}
\end{align}
Under the infinite-volume limit, the PBC spectrum for $tg\neq0$ is an ellipse in the complex plane.
Unlike the Hermitian tight-binding models, the OBC spectrum is drastically different from the PBC one.
By using a similarity transformation called $imaginary$ $gauge$ $transformation$, one can map the Hatano-Nelson model to a Hermitian matrix:
\begin{align}
H^{(\rm Herm)}&:= V_{r}^{-1} H^{\rm (HN)} V_{r}\notag\\
&=
\begin{pmatrix}
0&\sqrt{t^2-g^2}&\cdots\\
\sqrt{t^2-g^2}&0&\cdots\\
\vdots&\vdots&\ddots
\end{pmatrix},
\end{align}
where $[V_r]_{i,j}=r^i\delta_{ij}$ with $r=\sqrt{t+g/t-g}$.
We have assumed $t>g>0$ for simplicity.
Since a similarity transformation preserves the spectrum of any finite-dimensional matrix, the OBC spectrum of $H^{(\rm HN)}$ is given by that of $H^{(\rm Herm)}$, which is on the real axis owing to the Hermiticity.
Remarkably, the OBC spectrum is very different from the ellipse-like PBC spectrum.
Note that the OBC eigenstates of the Hatano-Nelson model are localized at one end because they are given by acting $V_r$ on the delocalized solutions of $H^{(\rm Herm)}$.

Actually, the spectral winding number of the PBC curve indicates the emergence of the the non-Hermitian skin effect.
Several observations about this correspondence were indicated in Refs. \cite{Gong-18, Lee-Thomale-19, Borgnia-19}, and related theorems were proven in Refs. \cite{OKSS-20,zhang-zhesen-chen}.
While the Hatano-Nelson model requires asymmetric hopping (or momentum-dependent dissipation), on-site dissipation can also induce the spectral winding number \cite{Yi-Yang-20}.
For example, the combination of the spin-momentum locked band and spin-dependent momentum-independent dissipation gives an effective momentum-dependence of the dissipation \cite{Okuma-Sato-21}. 
For applications in condensed matter physics, such an on-site dissipation may be more useful than the momentum-dependent dissipation.
This construction is analogous to an implementation of a topological superconductor in which the $p$-wave paring is effectively realized by the combination of the spin-momentum-locked band structure and the $s$-wave paring \cite{Sato-Ando,Sato-Fujimoto,Lutchyn,O-R-vO}.

\subsection{Boundary-dependent dynamical instability under non-Hermitian skin effect}
We here propose a dynamical instability that occurs under the PBC and does not occur under the OBC.
The non-Hermitian skin effect discussed in the previous subsection plays a central role in this proposal.
For simplicity, we consider the frequency-independent self-energy henceforth.

If anomalous terms are absent, there is no need to introduce the Nambu representation, and the bosonic Green's function is expressed in terms of a conventional form:
\begin{align}
    G^R(\omega)=\frac{1}{\omega-H-\Sigma^R}=:\frac{1}{\omega-H^{(\rm eff)}}.
\end{align}
In that case, however, $H^{(\rm eff)}$ cannot have eigenvalues with positive imaginary part, which means the absence of the dynamical instability.
Thus, the anomalous term is essential to observe the dynamical instability.
If the self-energy is negligibly small, there emerges the pseudo-Hermiticity.
In that case, the complex energy can be found only when the particle and hole bands experience the band crossing at the zero energy.
This means that one should give up the positive definiteness of $H^{(0)}_{\rm BdG}$ in order to observe the non-Hermitian skin effect because one cannot find a PBC spectral winding for the real spectrum.
Such an assumption is relevant in specific cases such as cold atom systems. See Ref. \cite{Mc-18,Flynn-20,BdG-non-bloch} for the dynamical instability in such cases.

For the relevance in various applications including solid-state physics, we consider the case where the particle and hole bands do not experience the band crossing.
For this sake, we use the combination of the Hatano-Nelson model (\ref{hnham}) and the anomalous terms.
Let us consider the following BdG Hamiltonian:
\begin{align}
    H^{(\rm eff)}_{\sigma\rm  BdG}=
    \begin{pmatrix}
    H^{(\rm HN)}-\alpha&i\Delta\\
    i\Delta&-[H^{(\rm HN)}-\alpha]^*
    \end{pmatrix},\label{hnbdg}
\end{align}
where $\Delta\in\mathbb{R}$ and $\alpha\in\mathbb{C}$ represent the anomalous term and a constant shift, respectively. This Hamiltonian is constructed to satisfy the particle-hole symmetry (\ref{omegadepphs}).
Note that the non-Hermiticity of the anomalous term originates from the Hermitian term of the $H^{(\rm eff)}_{\rm  BdG}$, and the essential non-Hermiticity is only in the normal term. 
Our motivation is to discuss the competition between these two types of non-Hermiticity.

\begin{figure}[t]
\begin{center}
 \includegraphics[width=8cm,angle=0,clip]{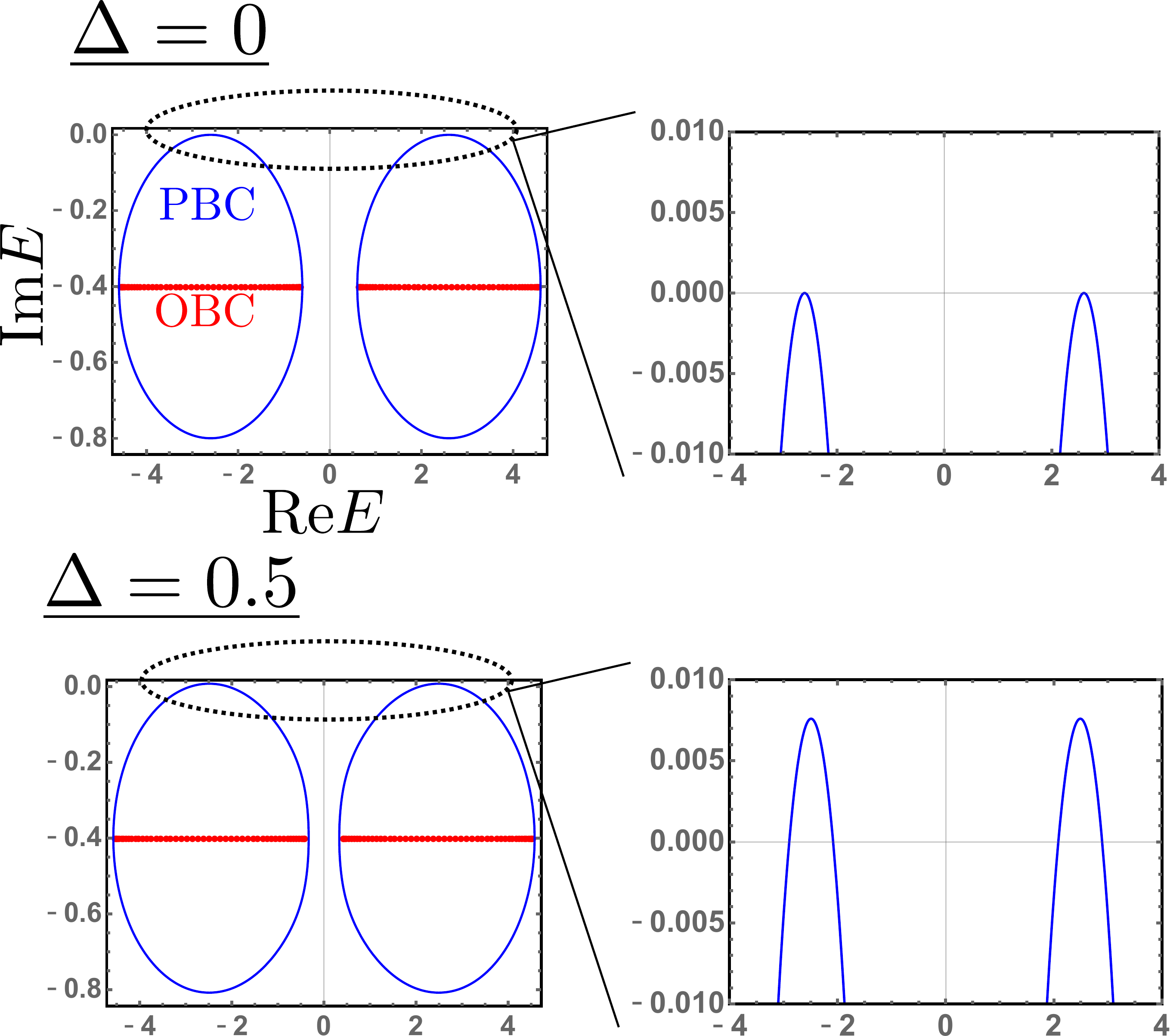}
 \caption{Bogoliubov spectra of the dissipative Bogoliubov-de Gennes Hamiltonian (\ref{hnbdg}) for $\Delta=0$ and $0.5$. The other model parameters are the followings: $t=1$, $g=0.2$, and $\alpha=-2(t+0.3)+2ig$. The PBC spectra are plotted by using the analytical expression (\ref{analyticalpbc}), while the OBC spectra are plotted by numerical diagonalization for size $L=50$.}
 \label{fig1}
\end{center}
\end{figure}

Under the PBC, the Bloch Hamiltonian for Eq.(\ref{hnbdg}) is given by 
\begin{align}
    H^{(\rm eff)}_{\sigma\rm  BdG}(k)=-i~\mathrm{Im}~\alpha+\left[E^{\rm (HN,PBC)}_{k}-\mathrm{Re}~\alpha\right]\sigma_z+i\Delta \sigma_x,
\end{align}
where $E^{\rm (HN,PBC)}_{k}$ is given in Eq.(\ref{pbcspectrum}).
Since the non-Hermiticity in the normal part should correspond to dissipation, we assume
\begin{align}
    \max_{k\in[0,2\pi)}\mathrm{Im}~ E_k^{\rm (HN,PBC)}=2g\leq \mathrm{Im}~\alpha.\label{causality}
\end{align}
In the absence of the anomalous term, the BdG Hamiltonian is reduced to the Hatano-Nelson model with a constant shift, and there is no dynamical instability owing to Eq. (\ref{causality}).
In the presence of the anomalous term, on the other hand, there is a possibility for the dynamical instability.
To see this, we consider the explicit form of the Bogoliubov spectrum:
\begin{align}
    E^{(\rm BdG)}_{k,\pm}&=-i~\mathrm{Im}~\alpha\pm\sqrt{\left[E^{\rm (HN,PBC)}_{k}-\mathrm{Re}~\alpha\right]^2-\Delta^2}\notag\\
    &=:-i~\mathrm{Im}~\alpha\pm\sqrt{A(|\Delta|)+i B},\label{analyticalpbc}
\end{align}
where $A,B\in\mathbb{R}$. 
The imaginary part of the square root is expressed as
\begin{align}
    \frac{\mathrm{Im}\sqrt{A+iB}}{B}=\frac{(A^2+B^2)^{\frac{1}{4}}}{\sqrt{(A+\sqrt{A^2+B^2})^2+B^2}}.
\end{align}
Since this form is a monotonically decreasing function of $A$ and $A(|\Delta|)$ is a monotonically decreasing function of $|\Delta|$, the absolute value of the square root is a  monotonically increasing function of $|\Delta|$.
Thus, the imaginary part of the Bogoliubov spectrum can be positive for sufficiently large $|\Delta|$, which leads to the dynamical instability.
Remarkably, one can choose a parameter so that the dynamical instability does not occur under the OBC because the OBC spectrum can be drastically different from the PBC one under the non-Hermitian skin effect.
In Fig.\ref{fig1}, we plot the PBC curve (\ref{analyticalpbc}) and the OBC curve that is obtained by a numerical diagonalization for $L=50$, where $L$ is the number of sites. 
We have set the largest imaginary part of PBC eigenvalues to be infinitesimally small value $0^{-}$ in the absence of the anomalous term.
Figure \ref{fig1} indicates that the introduction of the anomalous term leads to the dynamical instability under the PBC without the help of the band crossing, while the largest imaginary part of the OBC eigenvalues remains negative for finite $\Delta$ owing to the non-Hermitian skin effect, which means the absence of the dynamical instability.
Under the PBC, the complex energies with positive imaginary part are around $k\sim\pi/2$, which are the chiral modes (left movers) with finite real energies.
Thus, the collapse of the steady state is accompanied by a current circulating the PBC system.
If we set the largest imaginary part of PBC eigenvalues is a finite negative value, there is a threshold for $\Delta$ to induce the dynamical instability.
Note that too large $\Delta$ leads to the band crossing at the imaginary axis.
Note also that our proposal does not depend on the specific properties of the model and the situation.
One can discuss the boundary-dependent dynamical instability of the system with the non-Hermitian skin effect, which does not require the momentum-dependent dissipation as discussed in the previous subsection.
In addition, one can define the retarded Green's function both for equilibrium and non-equilibrium steady states, and the above discussion may be applicable to various systems.

\subsection{Dynamical instability related to infinitesimal instability}
In terms of the mathematics of the Toeplitz matrices,  a small $nonlocal$ perturbation that couples the different edges of a tight-binding model drastically changes the OBC spectrum \cite{Trefethen,Okuma-Sato-20}.
In particular, it becomes identical to the PBC curve with a spectral winding for an infinitesimally small perturbation after taking the infinite-volume limit.
In finite systems, the threshold for this spectral instability is taken to be an exponentially small value in terms of the system size.
In Ref. \cite{Okuma-19}, the author and a coworker discussed an infinitesimal instability against a $local$ perturbation of a spin-dependent non-Hermitian skin effect in which the direction of the asymmetric hopping depends on the spin.
The simplest model for this phenomenon is given by the stack of the Hatano-Nelson model:
\begin{align}
H_{\rm spin} \left( k \right)
= \left( \begin{array}{@{\,}cc@{\,}} 
	H^{(\rm HN)} \left( k \right) & 0 \\
	0 &  H^{(\rm HN)}  \left( -k \right) \\ 
	\end{array} \right). 
\end{align}
While the OBC spectrum of this model is on the real axis, the introduction of the transverse magnetic field that couples the different blocks of $H_{\rm spin}$ drastically changes the OBC spectrum.
In Ref. \cite{OKSS-20}, the author and coworkers used the notion of infinitesimal instability to characterize the $\mathbb{Z}_2$ skin effect protected by the time-reversal symmetry, which is characterized not by the spectral winding nubmer but by the $\mathbb{Z}_2$ topological invariant. 
In general, symmetry-protected skin effects should be unstable to symmetry-breaking perturbations.

The notion of infinitesimal instability would be relevant for the physics of dynamical instability.
If we introduce a small perturbation that couples the edges of Eq. (\ref{hnbdg}), we can expect a dynamical instability against it.
If we replace the normal part of Eq. (\ref{hnbdg}) with the system with a symmetry-protected skin effect, we can expect a dynamical instability due to a small symmetry-breaking local perturbation under the OBC.
Note that a non-dissipative Bogoliubov-de Gennes Hamiltonian treated in Ref. \cite{BdG-non-bloch} shows an infinitesimal instability, which was also discussed in terms of a dynamical instability.

\section{Discussion\label{discussion}}
We remark several points about the band crossing between particle and hole bands and the Landau instability.
Strictly speaking, the absence of the band crossing does not directly mean the absence of the Landau instability.
While we have assumed the frequency-independent dissipation, it means that the density of states for the particle bands calculated by $D_{\rm part}(\omega)=1/\pi\mathrm{Im}\sum_{i:\mathrm{particle}} (\omega-E_i)^{-1}$ , where $E_i$'s are eigenvalues of $H^{(\rm eff)}_{\sigma\rm  BdG}$, takes a finite value at $\omega<0$, which indicates the presence of the Landau instability. In this sense, $\omega$-dependence of the dissipative BdG Hamiltonian should be considered for more realistic calculations.
For this sake, one should treat $\omega$-dependent equation (\ref{poles}) to get the Bogoliubov spectrum.
Let $z_0$ be a Bogoliubov eigenvalue with positive imaginary part of some frequency-independent $H^{(\rm eff)}_{\sigma\rm  BdG}$.
From this Hamiltonian, we can construct a frequency-dependent Hamiltonian with the dynamical instability and without the Landau instability by choosing $\Sigma^R(\omega)$ so that $\Sigma^R(z_0)=\sigma_z(H^{(\rm eff)}_{\sigma\rm  BdG}-H^{(0)}_{\sigma\rm  BdG})$ and $D_{\rm part}(\omega<0)=0$, at least mathematically.
Thus, the essence of our result does not suffer from the Landau instability.

We also mention the relationship between the non-Hermitian skin effect and the Landau instability.
Although one might expect that the skin effect can also induce the sensitivity of the Landau instability to the boundary condition, such an expectation may be incorrect.
If the smallest real part of the PBC curve is negative and that of the OBC curve is positive, the Landau instability is induced for both cases.
This is because the relevant quantity for the Landau instability may not be the Bogoliubov complex energy but the density of states around zero energy.
As shown in Ref. \cite{Okuma-correlated},
the density of states is insensitive to the non-Hermitian skin effect in the absence of complex eigenvalues with positive imaginary part, which implies the insensitivity of the Landau instability to the boundary condition.

\acknowledgements
I thank Masatoshi Sato for a fruitful discussion about free bosonic BdG Hamiltonians.
I also thank Terufumi Yamaguchi for a fruitful discussion about Green's functions.
This work was supported by JST CREST Grant No.~JPMJCR19T2, Japan. N.O. was supported by JSPS KAKENHI Grant No.~JP20K14373.

\bibliography{NH-topo}
\end{document}